\pdfoutput=1
\documentclass[prl,reprint,twocolumn, aps, amsmath, amssymb,floatfix]{revtex4-1}
\usepackage{graphicx,amsmath,amssymb}
\usepackage[usenames]{color}
\usepackage{bm}
\begin{document}

\title{Landau-level-mixing and the ground state of the $\nu=5/2$ quantum Hall effect}
\author{Edward H. Rezayi}
\affiliation{Department of Physics, California State University Los Angeles, Los Angeles, CA 90032, USA}

\begin{abstract}
Inter-Landau-level transitions break particle
hole symmetry and will choose either the Pfaffian or the anti-Pfaffian state
as the absolute ground state at 5/2 filling of the fractional quantum Hall effect. 
An approach based on truncating the Hilbert space has favored  the anti-Pfaffian. A second approach based
on an effective Hamiltonian produced the Pfaffian. In this letter
perturbation theory is applied to finite sizes without bias to any specific 
pseudo-potential component. This method also singles out the anti-Pfaffian. 
A critical piece of the effective Hamiltonian, which was absent in previous studies, 
reverts the ground state at 5/2 to the anti-Pfaffian.

\end{abstract}

\maketitle
 Topological phases of matter 
have been the focus of much recent theoretical interest. 
A prime example of such a phase\cite{Wen-Top-Order}, which is realized in experiment, is the 
fractional quantum Hall effect\cite{Tsui-etal,Laughlin} (FQHE). 
The most intriguing of these quantum Hall states 
occurs in the half-filled first excited Landau level (1LL)\cite{Willett-even}.  Breaking the 
odd-denominator trend of the lowest Landau level (LLL), the state at
$\nu=5/2$ is one of the strongest FQHE states in the 1LL\cite{Pan-etal-5-2,Xia-etal}.
 It has been the subject of numerous studies and is widely believed 
to be in the universality class of the  Moore-Read (MR) state\cite{MR}, which is a fully polarized 
$p_x+ip_y$ paired state\cite{Read-Green} of composite fermions\cite{Jain}.  
The MR state possesses richer excitations than states dominating the LLL. 
In particular, the presence of the neutral Majorana fermion mode 
 results in quasi-particle excitations exhibiting non-Abelian statistics\cite{MR,Nayak-Wilczek,RR-MRQH,Read-p+ip}.
This property has greatly increased interest in the 5/2 effect, in part due to its potential 
for quantum information processing\cite{Kitaev-anyons,Nayak-etal-RMP}.  In this letter we will
assume the MR state to be the underlying phase of the 5/2 effect.
\cite{Morf,Rezayi-Haldane-5/2, XinWan-etal,Sheng-Haldane-DMRG}

{\it Particle-hole symmetry and the anti-Pfaffian-} 
Earlier studies of the 5/2 state disallowed virtual  inter-LL transitions, 
 which are only exact at infinite cyclotron energy $\hbar\omega$.
In that case, because of particle-hole (P-H) symmetry,
both the MR or (Pfaffian) state and its P-H 
conjugate\cite{Levin-Halperin-Rosenow,Nayak-Fisher} (the anti-Pfaffian)
 are equally valid candidates for the 5/2 state. However, 
while both states display non-Abelian statistics,  they are distinct topological 
phases\cite{Levin-Halperin-Rosenow,Nayak-Fisher} of matter. The system will then be forced to 
choose one 
by spontaneously breaking P-H symmetry.  On the other hand, the ubiquity of LL-mixing provides a 
P-H symmetry breaking field. A measure of the mixing strength is 
given by the ratio of typical Coulomb energy $\varepsilon=e^2/4\pi\epsilon_0\ell$
to the cyclotron energy $\kappa=\varepsilon/\hbar\omega$, where $\ell$ is 
the 
magnetic length.
In  experiments, $\kappa$ varies from 0.8 to 2.8 and 
either the Pfaffian (Pf) or the anti-Pfaffian (aPf) will be favored as the ground state.  
Until now, however, which one has not been definitively determined.

To address the effects of P-H symmetry-breaking, several studies have been carried 
out using different approximations.  Simon 
and the present author\cite{Rezayi-Simon-LLmix} employed a truncated model of LL-mixing 
by keeping only 3 LLs (3-LL model). Additionally we controlled the number of particle and 
hole excitations that result from inter-LL transitions. 
In a subsequent work by Zaletel {\it et. al.}\cite{Zaletel-etal} this last restriction was completely removed.  The
authors employed an infinite density matrix renormalization group (iDMRG)\cite{Zaletel-Mong-Pollmann}.  The method incorporates the
matrix product form\cite{Zaletel-Mong, Bernevig-etal} of FQHE states
on a cylinder of infinite length but finite radius.  
The iDMRG also
kept a finite number of LLs (3-5), but could reach cylinder circumferences of 20 magnetic lengths.  
A square torus of this dimension would include about 30 electrons, which is a considerable
increase in system size over previous studies.
Both approaches predict the aPf to be favored irrespective of the strength of $\kappa$.  

On the other hand, Wojs {\it et. al.}\cite{Wojs-Toke-Jain,*[{This work was based on the results of
 Bishara and Nayak, which did not include the $M=9$ 3-body pseudo-potentials; the two-body
pseudo-potentials were subsequently revised in Ref.\cite{Peterson-Nayak} by normal 
ordering of the 3-body interactions. Interestingly, the overlap of the ground state with the aPf
only improves if $H(M=9)$ is added to the Bishara-Nayak Hamiltonian; }] Bishara-Nayak} 
and more recently Pakrouski {\it et. al.}\cite{Pakrouski-etal} have concluded that the Pf is the ground
state. The authors used 
the effective 2-body\cite{Haldane-2body} and 3-body\cite{Simon-Rezayi-Cooper} 
pseudo-potentials\cite{Simon-Rezayi,Peterson-Nayak,Sodemann-MacDonald} that 
include LL-mixing corrections to the lowest order in $\kappa$. At first sight, these calculations have cast doubt on the results of the 3-LL model. 

In this letter the issue is revisited in order to resolve this discrepancy and 
to compare the 3-LL model predictions to perturbation theory results.
The recent calculations of infinite-size pseudo-potentials produced a 3-way 
agreement\cite{Simon-Rezayi,Peterson-Nayak,Sodemann-MacDonald} on their values
and are not the cause of the discrepancy.  

Our approach to calculating the effective Hamiltonian
closely follows the previous methods, 
but uses the torus geometry for finite-size  systems.
This is the only compact geometry that avoids the shift\footnote{The shift S is defined by 
the flux-charge equation $N_\phi=\nu_v^{-1} N_e-S$, where $\nu_v$ is the filling factor of the 
valence electrons.  On the sphere S=3 and -1 for the Pf and aPf respectively.} and facilitates the
comparison between Pfaffian and anti-Pfaffian.  A study on the sphere is also
reported in this letter which corroborates the torus results.  For this and subsequent 
torus results we will follow the method
of Wojs {\it et. al.} and Pakrouski {\it et. al.} and use infinite-size perturbation 
theory (PT) values of the 2-body and the 3-body 
pseudo-potentials.  

The Hamiltonian allowing inter-LL transitions is 
\begin{equation}
{\cal{H}}=\frac{1}{2}\sum_{\{m_i\}} \langle m_1\ m_2\lvert V \rvert m_3\ m_4\rangle C^\dagger_{m_1} C^\dagger_{m_2}C_{m_3}C_{m_4},
\end{equation} 
where $m$ is the combined label of LL index $n$, linear momentum $k$, and spin. 
$V(r)$ is the Coulomb interaction.  We will consider the electron layer to have zero width.
The matrix elements in {\cal{H}}, which are independent of spin,  are calculated 
self-consistently for up to 31 
LLs for each size that we studied.  Another important difference with 
infinite-size calculations is that the  
3-body pseudo-potential corrections are not singled out by their relative angular
momentum but are automatically included in their entirety.  

The effective Hamiltonian, which includes the lowest order LL-mixing corrections can be written as:

\begin{eqnarray}
H_{\text{eff}}= H_1 +\kappa\sum^\prime_p 
{\frac{{\cal{H}}\lvert p \rangle\langle p \lvert 
{\cal{H}}} {E_0-E_p}},
\label{eq:SH}
\end{eqnarray}
where $H_1$ is the 
Hamiltonian for electrons in the partially filled 1LL and $\lvert p\rangle$ is 
an intermediate state with kinetic energy $E_p$, which can have at most two electrons
in the excited LLs with index $n>1$. $E_0$ is the kinetic energy of the appropriate  basis
states (described below) and is
dropped from $H_{\text{eff}}$.  The prime on the sum restricts $p$ so that $E_p \ne  E_0$. 
In Eq. \ref{eq:SH}, the Hamiltonians are expressed in units of $\varepsilon$ and kinetic 
energies are given in units of the cyclotron energy.

We will generally consider the matrix elements $\langle i \lvert H_{\text{eff}} \rvert j\rangle $,
where the set \{$\lvert i\rangle$\} is a relevant Slater determinant basis for $N_e$ valance
electrons in the 1LL and includes a lowest LL filled with both spins. We will choose $N_e=2$
or 3 when calculating pseudo-potentials. 
This form is also readily applicable when the degeneracy
is not completely lifted after the action of $H_1$, which is the case for certain odd 
electron numbers on a hexagonal torus\cite{Zlatko-QW}. In this case, the
ground state that represents the basis \{$\lvert i \rangle$\} is a doublet.
Extra degeneracies of this type 
result from the combination of anti-unitary P-H conjugation and unitary discrete rotational symmetries\cite{Tinkham}.

It can be seen  from the definition of the basis states that the two creation
operators of the right ${\cal{H}}$ acting on $\lvert i\rangle$ 
(and destruction operators of the left ${\cal{H}}$ acting on 
$\langle j\rvert$) will have LL indices of either 0 or  1.
It is therefore convenient to organize ${\cal{H}}$ in terms of the number of LLL holes (0,1,2) in
$\lvert p\rangle$. For example, in the case of two holes, one can ignore the intermediate states
$\lvert n_1,n_2\rangle$, where $n_i>1$ are excited LL indices, since these terms do not
depend on the specifics of $\lvert i \rangle$ and $\lvert j\rangle$ and amount to 
an overall constant in Eq. \ref{eq:SH}.  The last term of $H_{\text{eff}}$ can be represented by Feynman diagrams,
which  have already appeared in print\cite{Peterson-Nayak,Sodemann-MacDonald} and will not be 
repeated here. It is then
a bookkeeping exercise; some relevant details can be found in the appendices 
B and C of
Sodemann and MacDonald. The results are 1, 2, and 3-body effective interactions for 
electrons in the 1LL. 
However, the one-body potentials, by translational symmetry, are independent of orbital index. Since they only serve to 
modify the chemical potential, they will also be ignored. 
It will be assumed throughout that the states in the 1LL are fully spin-polarized, 
which is consistent with recent experiments\cite{Eisenstein-spin-diode,Gamez-etal} as 
well as 
with previous calculations\cite{Morf,Feiguin-etal,Rezayi-Simon-LLmix}. However,
virtual reversed spin excitations from the LLL are 
allowed. Even these make a negligible difference and can
be ignored. 

\begin{figure}[t]\centering
         \includegraphics[width=.45\textwidth]{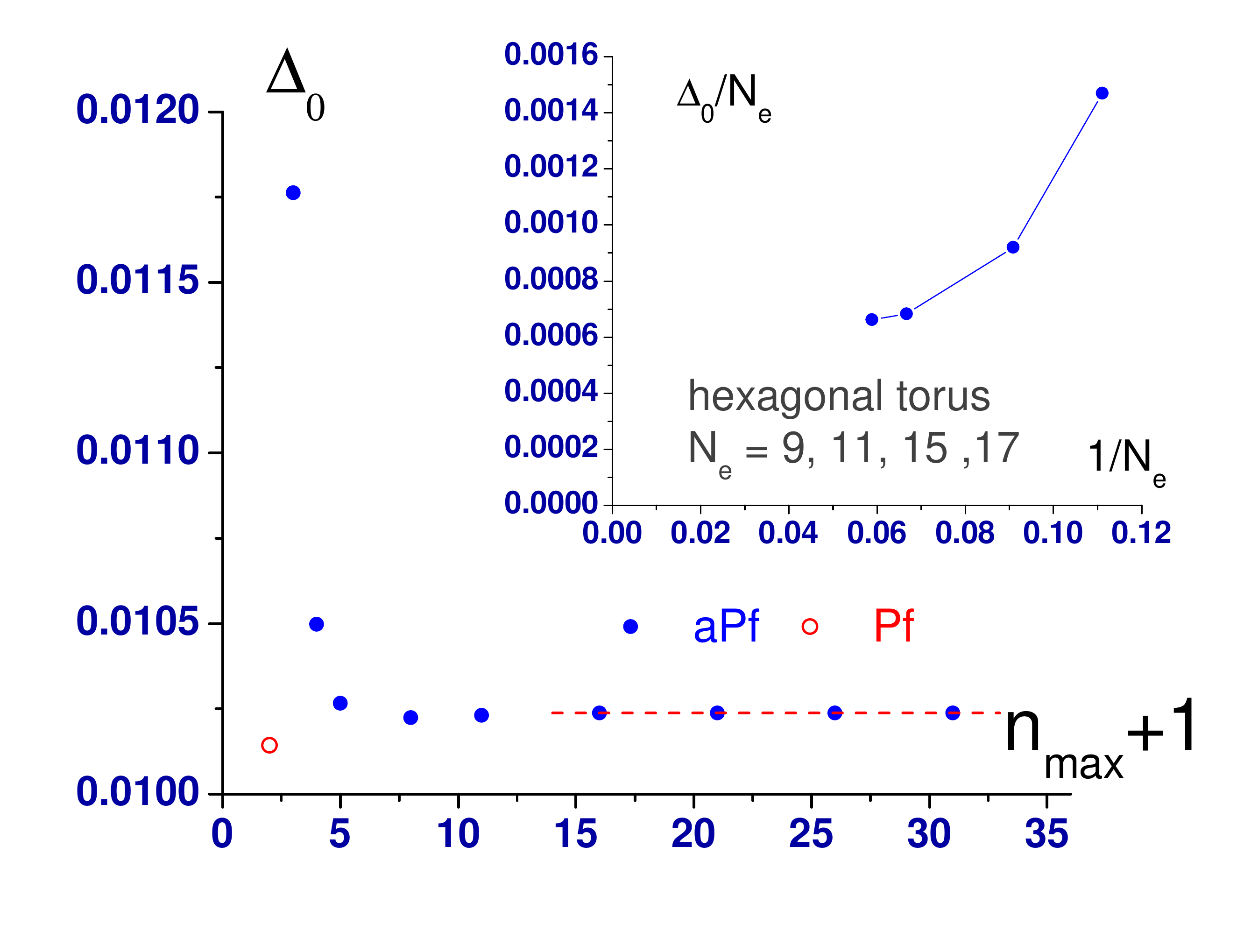}
         \caption{ (Color online) Plot of Pf-aPf energy difference 
$\Delta_0=E_{\text{Pf}}-E_{\text{aPf}}$ (in units of $\varepsilon\kappa$) between the aPf 
and Pf as a function of the number of 
LLs included in the present calculation ($n_{\text{max}}$ is the index of the highest LL). The
Pf ground state occurs only when the lowest two LLs are kept.
 The inset shows $\Delta_0$ for odd sizes divided by the number of electrons plotted vs $1/N_e$.}
        \label{fig:gap0vsLL}
      \end{figure}
In the case of a doublet ground state of the $H_1$, we follow Landau and Lifshitz\cite{Lan-Lif} 
and diagonalize
the LL-mixing part of the $H$ in the 2-dimensional Hilbert space spanned by the doublet, 
which separates the Pf and the aPf components without mixing-in any other state of the $H_1$.
Fig. \ref{fig:gap0vsLL} shows the difference in the ground state energy 
difference $E_{\text{Pf}}-E_{\text{aPf}}=\Delta_0$  as  a 
function of the number of LLs included in the sum of Eq. \ref{eq:SH}.  As observed in the 3-LL model, the aPf is favored 
unless only the lowest 2 LLs are kept.  The  inset shows $\Delta_0/N_e$ as a function of $1/N_e$.
While the linear regime has not quite been reached and it is difficult to extrapolate to large sizes, 
the results are consistent with a ``gap'' that is extensive\cite{Zaletel-etal}.  

\begin{figure}[t]\centering
         \includegraphics[width=.45\textwidth]{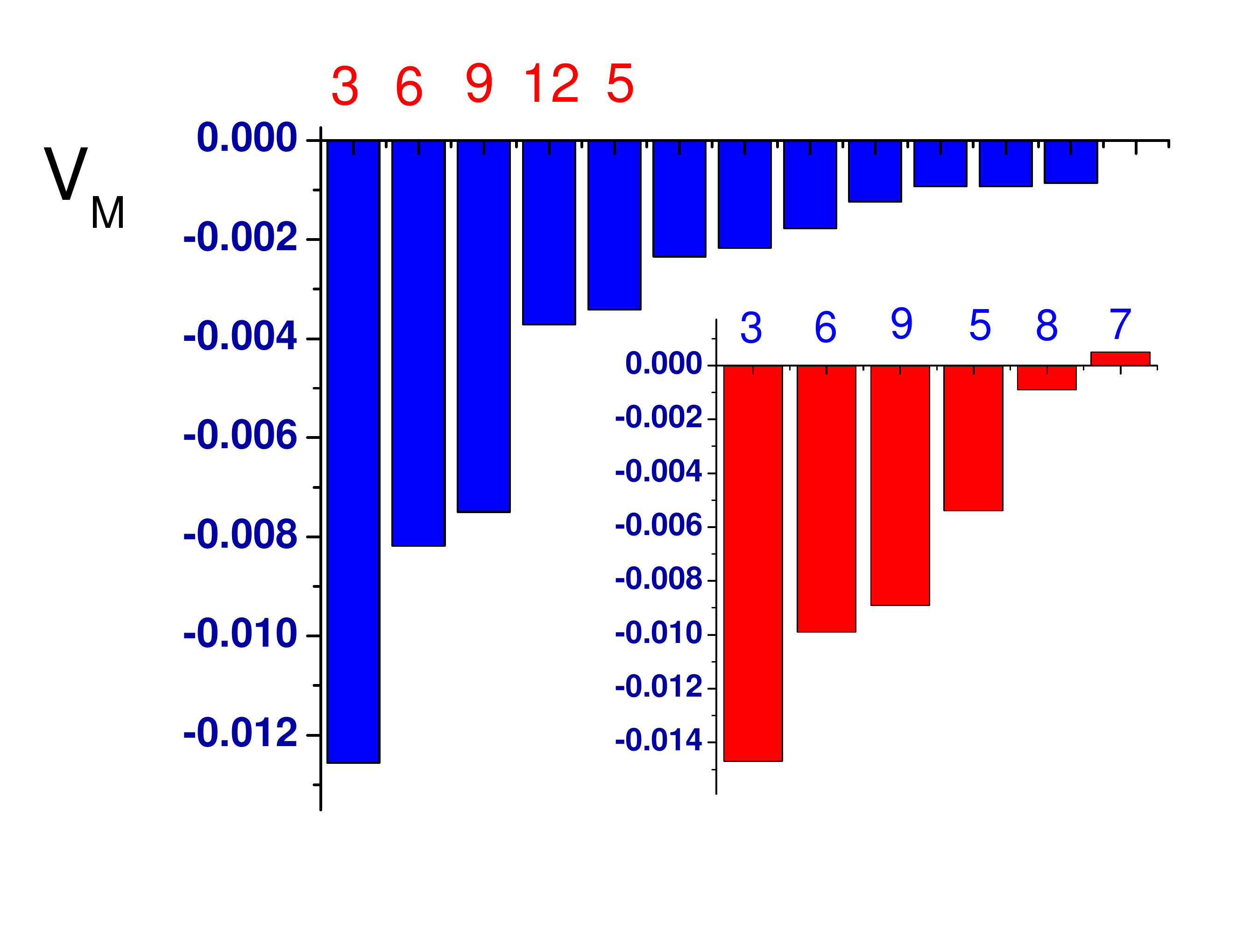}
         \caption{ (Color online) The 3-body pseudo-potentials (in units of $\varepsilon\kappa$) obtained on the torus 
for $N_\phi =30$ and arranged in ascending order, where $N_\phi$ is the number of magnetic 
flux quanta through the system. The relative angular momentum of the 5 lowest potentials are
indicated above the corresponding bar. The inset shows the infinite-size PT values. In contrast,
the pseudo-potentials for the first point in Fig. \ref{fig:gap0vsLL} (a 2-LL system  
that results in the Pf ground state) are repulsive: 0.0148 $(M=3)$, 0.0085 (6), 0.0029 (8), 
0.0023 (9), etc.
                        }
        \label{fig:VMhist}
      \end{figure}

{\it 2-body and 3-body pseudo-potentials.-} A very practical  way of presenting the electronic 
interaction potential for quantum Hall states is to express them in terms of the energy (or pseudo-potential) for 
$n$-electrons in a state of fixed relative angular momentum.  It was first introduced by 
Haldane\cite{Haldane-2body} for the 2-body case and later extended to the $n$-body case by 
Simon {\it et. al.}\cite{Simon-Rezayi-Cooper}.
Fig. \ref{fig:VMhist}  gives the values of the 3-body pseudo-potentials 
$V_M$ (in units of $\varepsilon\kappa$),  
in an ascending order. The first 5 are labeled by their relative angular momenta $M$. 
The inset gives the results of Sodemann
and MacDonald, who crucially calculated $V_9$. In both cases 
the most important pseudo-potentials have relative angular momenta that are 
multiples of 3\footnote{It can be seen that finite-size effects become larger as $M$
is increased. For larger $M$s the average relative distance between the particles
increases and may become comparable to the linear dimension of the system. 
As a result the $M=12$ may be off by 15-20\% of its infinite-size value}.
 The main difference between the present effective Hamitonian and 
those of Wojs {\it et. al. }and Pakrouski {\it et. al.} is the $M$=9 3-body pseudo-potential.
It will be shown that $V_9$ 
has a critical role in determining the nature of the ground state.
If included, the ground state reverts to the aPf. 
Table \ref{tab:V3M} gives the differences of $V_M$ from the $M=3$ value for finite-size systems in the present 
study and compares them with their infinite-size values.
Table \ref{tab:v2m} is the same as Table \ref{tab:V3M} but for 2-body pseudo-potentials, which already are
very close to their infinite size values. 

\begin{table}
 \begin{tabular}{r|c|c|c|c}
$N_\phi $&$ V_6- V_3$&$V_9-V_3$&$ V_{12}-V_3$&$ V_5- V_3$\\ \hline\hline
30&0.0044 (8.3\%)&0.0051 (12\%)&0.0088 &0.0091 (2.2\%)\\
32&0.0044 (8.3\%)&0.0051 (12\%)&0.0090 &0.0092 (1.1\%)\\
34&0.0045 (7.0\%)&0.0052 (10\%)&0.0090 &0.0092 (1.1\%)\\
$\infty$&0.0048 &0.0058 &NA&0.0093
 \end{tabular}
\caption{Comparison of the differences between 3-body pseudo-potentials $V_M$ for polarized
electrons and finite flux sizes and their infinite size counterparts. The percent differences are given 
in the parenthesis.}
\label{tab:V3M}
\end{table}
\begingroup
\squeezetable
\begin{table}
\begin{tabular}{r|c|c|c|c}
$N_\phi $&$\delta v_3-\delta v_1$&$\delta v_5-\delta v_1$&$\delta v_7-\delta v_1$&$\delta v_9-\delta v_1$\\ \hline\hline
30&0.1094 (0.45\%)&0.1767 (0.62\%)&0.1994 (0.75\%)&0.2067 (0.86\%)\\
32&0.1094 (0.45\%)&0.1769 (0.51\%)&0.1996 (0.65\%)&0.2069 (0.77\%)\\
34&0.1095 (0.36\%)&0.1770 (0.45\%)&0.1998 (0.55\%)&0.2071 (0.67\%)\\
$\infty$&0.1099 &0.1778 &0.2009 &0.2085
\end{tabular}
\caption{The LL-mixing corrections $\delta v_m$ for 2-body pseudo-potentials $v_m$
(in units of $\varepsilon\kappa$) 
relative to their respective $\delta v_1$ values and for 3 different flux sizes on the torus. 
For polarized electrons only odd values of relative angular momentum $m$ are relevant.
The
percent differences from infinite PT results (last row) are given in parenthesis.} 
\label{tab:v2m}
\end{table}
\endgroup

In the remainder of this letter the effective Hamiltonian for the infinite system
(see Table \ref{tab:v2m} and the inset of Fig. \ref{fig:VMhist}) will be used. 
While the first five pseudo-potentials 
(in order of their relative angular momenta)
are unique, the $M=9$ has two states and a choice of basis is necessary. 
The Hamiltonian matrix for $M=9$ and its corresponding basis wave functions are described 
by Sodemann and MacDonald\cite{Sodemann-MacDonald} and Laughlin\cite{Laughlin-3part} respectively. It is: 

\begin{eqnarray}
H(M=9)\ =&\ -0.0088\bm{\lvert 0,3\rangle\langle 0,3\rvert}+0.0033\bm{\lvert 3,1\rangle\langle 3,1\rvert} & \nonumber \\*
&+0.0007\bm{\left [\lvert 0,3\rangle\langle 3,1\rvert+ \lvert 3,1\rangle\langle 
0,3\rvert}\right],&
\label{eq:SMwf}
\end{eqnarray} 
where the states $\bm{\lvert l,m \rangle}$  have relative angular momentum  $M=2l+3m$. The relevant 
pseudo-potentials and their projection operators are
obtained by finding the eigenvalues and eigenvectors of Eq. \ref{eq:SMwf}. However, $H(M=9)$
is well approximated by just the first term, which is the most dominant by far.  
\begin{figure}[t]\centering
         \includegraphics[width=.45\textwidth]{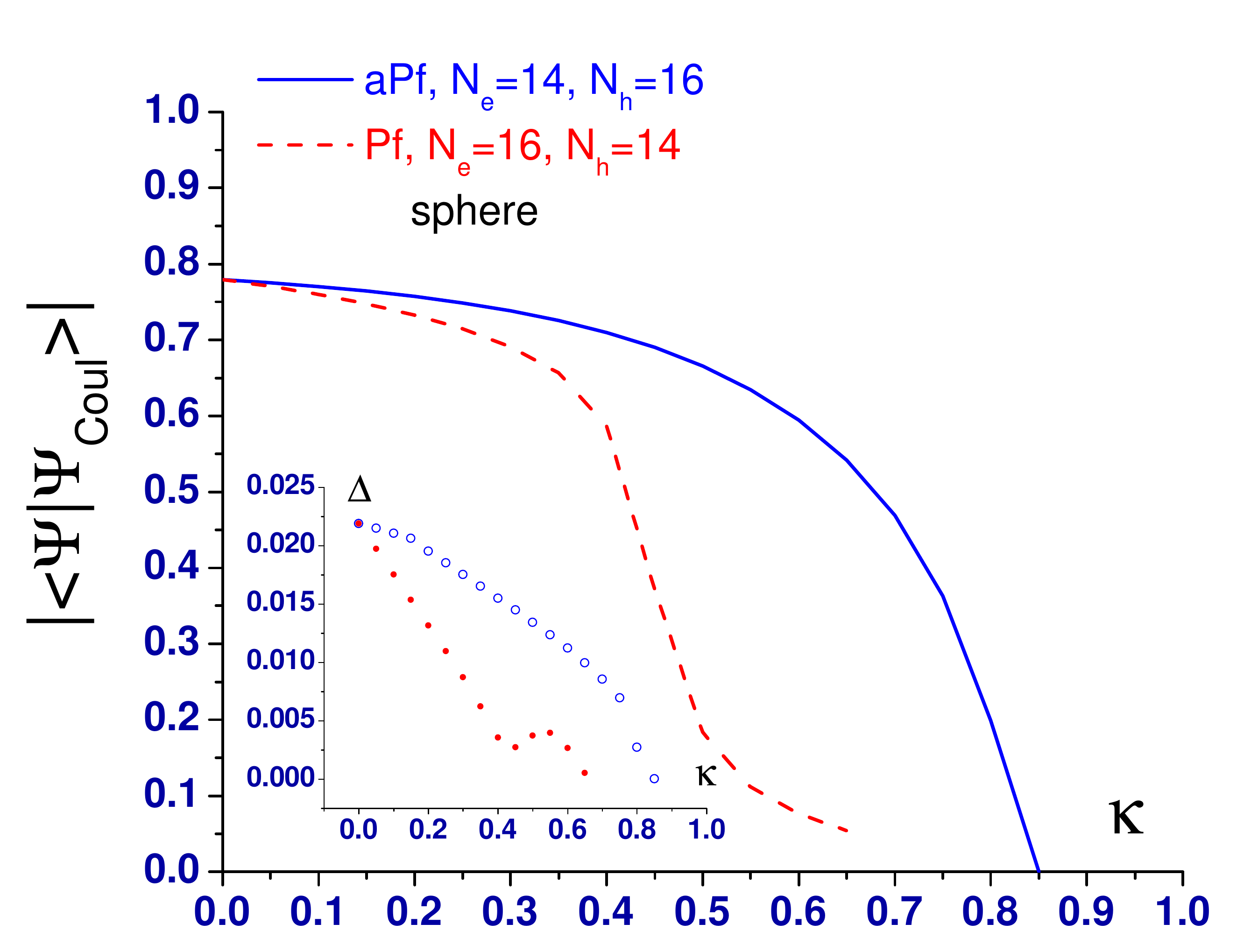}
         \caption{(Color online) Plot of overlaps of the ground state of the coulomb interaction with the aPf(solid) and Pf(dashed) 
vs. $\kappa$. The inset is the respective energy gap (in units of $\varepsilon$)  
vs. $\kappa$ in each case. The sudden bends and kinks in the gaps result from level crossings 
of the lowest excited state.
                        }
        \label{fig:SPOver}
      \end{figure}

The effective Hamitonian, including the $M=9$ term, is then diagonalized. In general, the eigenvalues 
and eigenvectors will have a non-linear dependence on $\kappa$.
 However, the linear 
regime will suffice for our conclusions.

The majority of numerical calculations on the 5/2 state have been carried out in the spherical 
geometry\cite{Haldane-2body}
 Because of the different shifts for Pf and aPf on the sphere, P-H symmetry is not 
applicable. Comparing the Pf with $N^P_e$ electrons to the aPf with an equal number of holes
$N^A_h=N^P_e$, forces the number of electrons to be related by $N^A_e=N^P_e-2$.
This difference in system sizes excludes any meaningful energetic comparison between Pf and aPf.

As in the Pakrouski {\it et. al.} paper, we compare overlaps and gaps between the ground and the
first excited state.  As noted by these authors, the gap in the spectrum 
tracks the respective overlaps and may provide another diagnostic tool. 
Fig. \ref{fig:SPOver} shows the results for $N_e=16$ for the Pf and $N_e=14$ for
the aPf.  It can be seen that now the aPf has the larger overlaps and gaps. 

\begin{figure}[t]\centering
         \includegraphics[width=.45\textwidth]{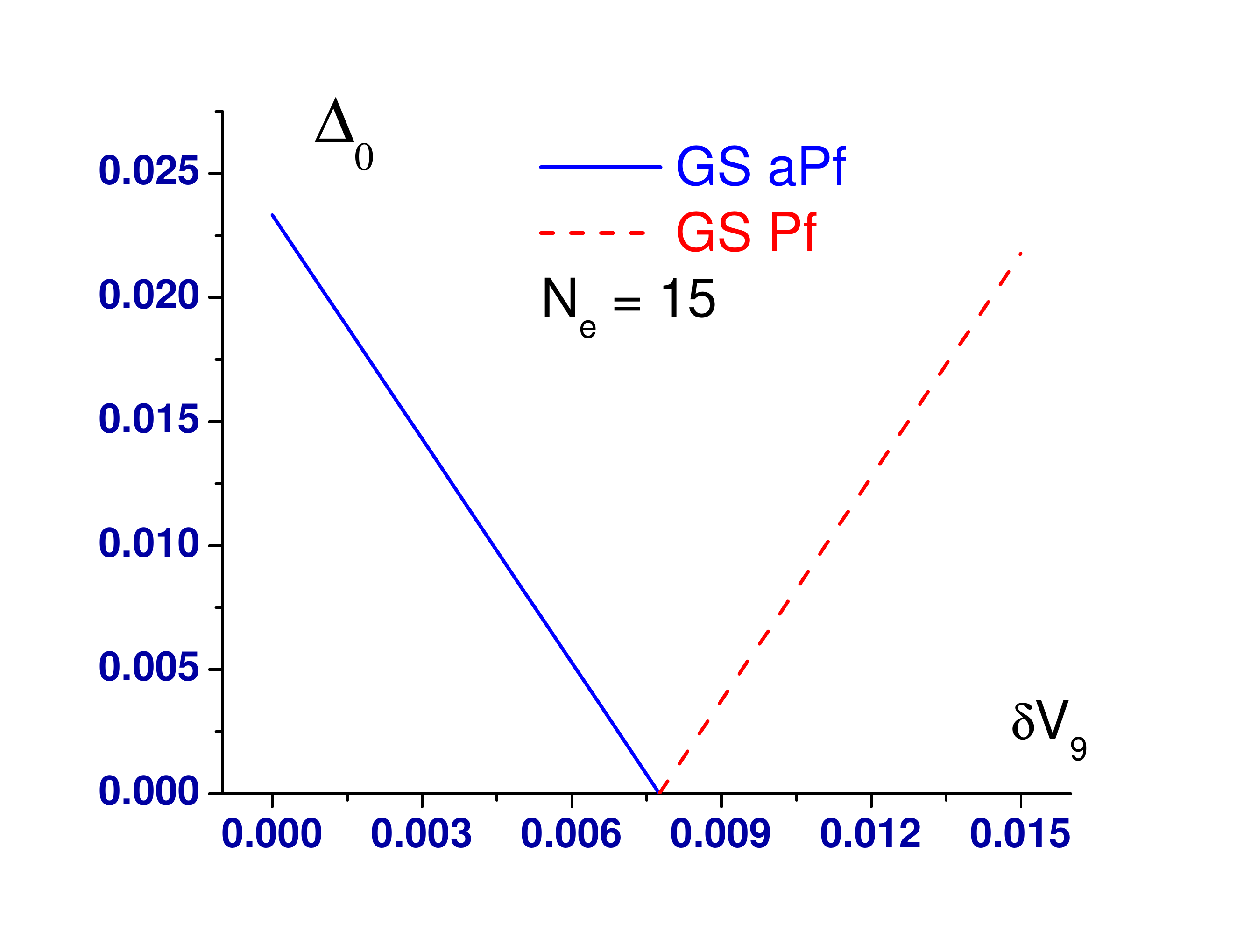} 
         \caption{(Color online) Plot of the aPf-Pf energy difference for $N_e=15$ as a function of $\delta V_9$.  The solid line 
indicates the ground state is the aPf, while the dashed line indicates the Pf ground 
state. The transition point is at $\delta V_9 \approx 0.0077$. Before the transition the 
overlap of the ground state with the aPf is 0.70; there is no overlap with the Pf.  After the 
transition these values are exchanged.
                        }
        \label{fig:gapvsv9}
      \end{figure}

{\it Transition to the Pfaffian.-} Clearly, there will be a phase transition as the $V_9$ 
pseudo-potential  approaches zero.  Making it less attractive weakens 
the aPf; at $V_9\approx 
-0.0011$ a first order transition to the Pf is observed. The results for the $N_e=15$ doublet on the torus 
is shown in Fig. \ref{fig:gapvsv9}. In this case the aPf-Pf energy difference  $\Delta_0$ 
displays a perfectly linear dependence on $V_9\ (=-0.0088+\delta V_9)$.
It is noteworthy that to reach the point of transition the magnitude of $V_9$ 
has to be reduced by approximately 90\% of its nominal value, which is a measure of 
how robust the ground state is. While the effect of a finite thickness of the 2-D layer 
has been ignored here, it seems unlikely that it could have a qualitative effect on the 
properties of the ground state.

\begin{figure}[t]\centering
         \includegraphics[width=.45\textwidth]{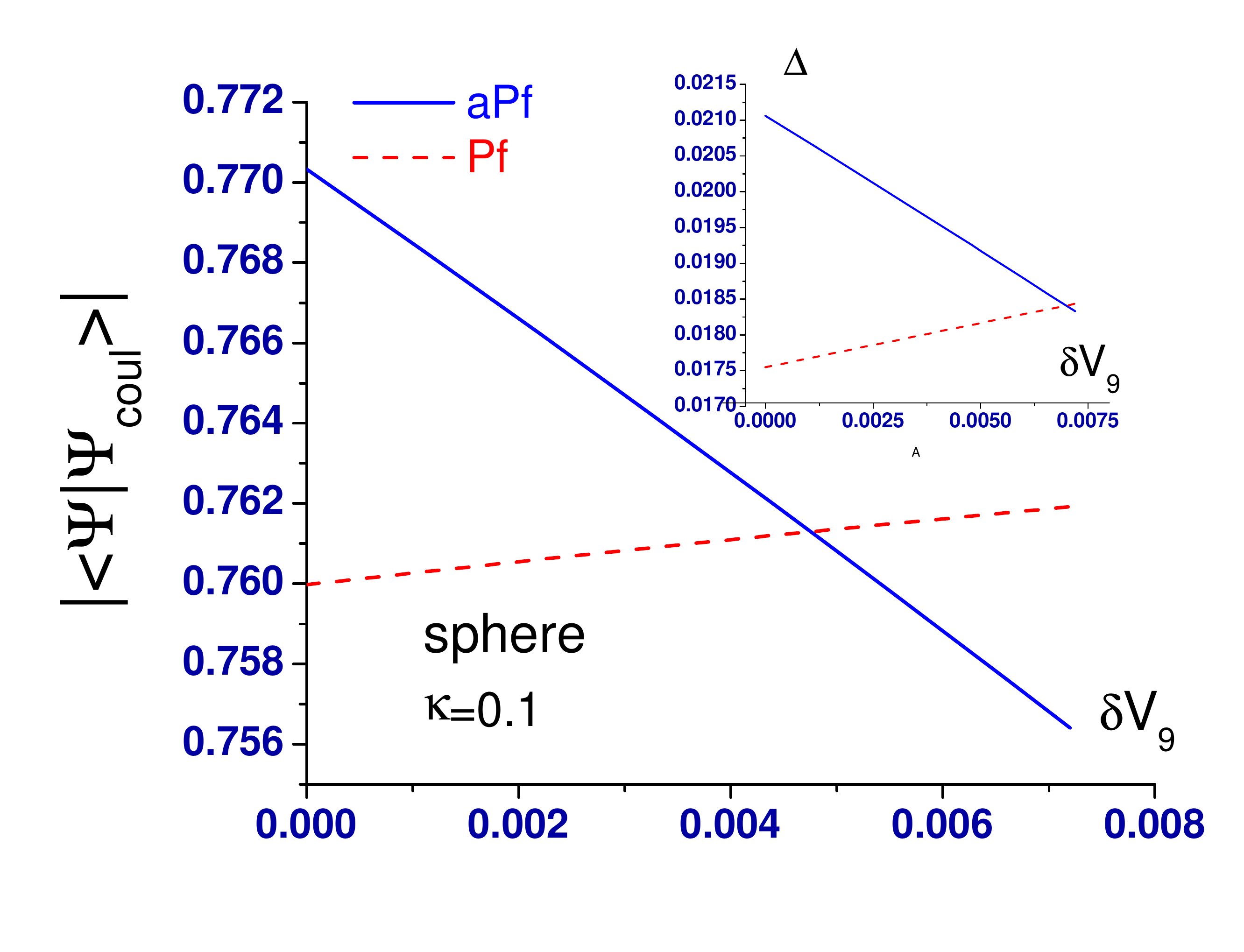}
         \caption{ (Color online) Plot of overlaps and respective gaps (inset in units of $\varepsilon$ ) vs. $\delta V_9$
of the aPf ($N_e=14$) and the Pf ($N_e=16$) on the sphere for $\kappa=0.1$. 
                        }
        \label{fig:spheregapvsv9}
      \end{figure}

The linear dependence on $V_9$ is also seen on the sphere for both gaps and overlaps 
(Fig. \ref{fig:spheregapvsv9}).  The trend of
the gaps, shown in the  inset, track those of the corresponding overlaps. The Pf and the aPf 
have opposing dependence on $V_9$.  It is difficult to pin-point the transition point 
with any accuracy here, but it is somewhere between
the crossing points of overlaps and the gaps. This is just a finite-size effect due to the size
mismatch between the  Pf and aPf, which prevents the two states from competing for the 
same ground state.

\begin{figure}[b]\centering
         \includegraphics[width=.45\textwidth]{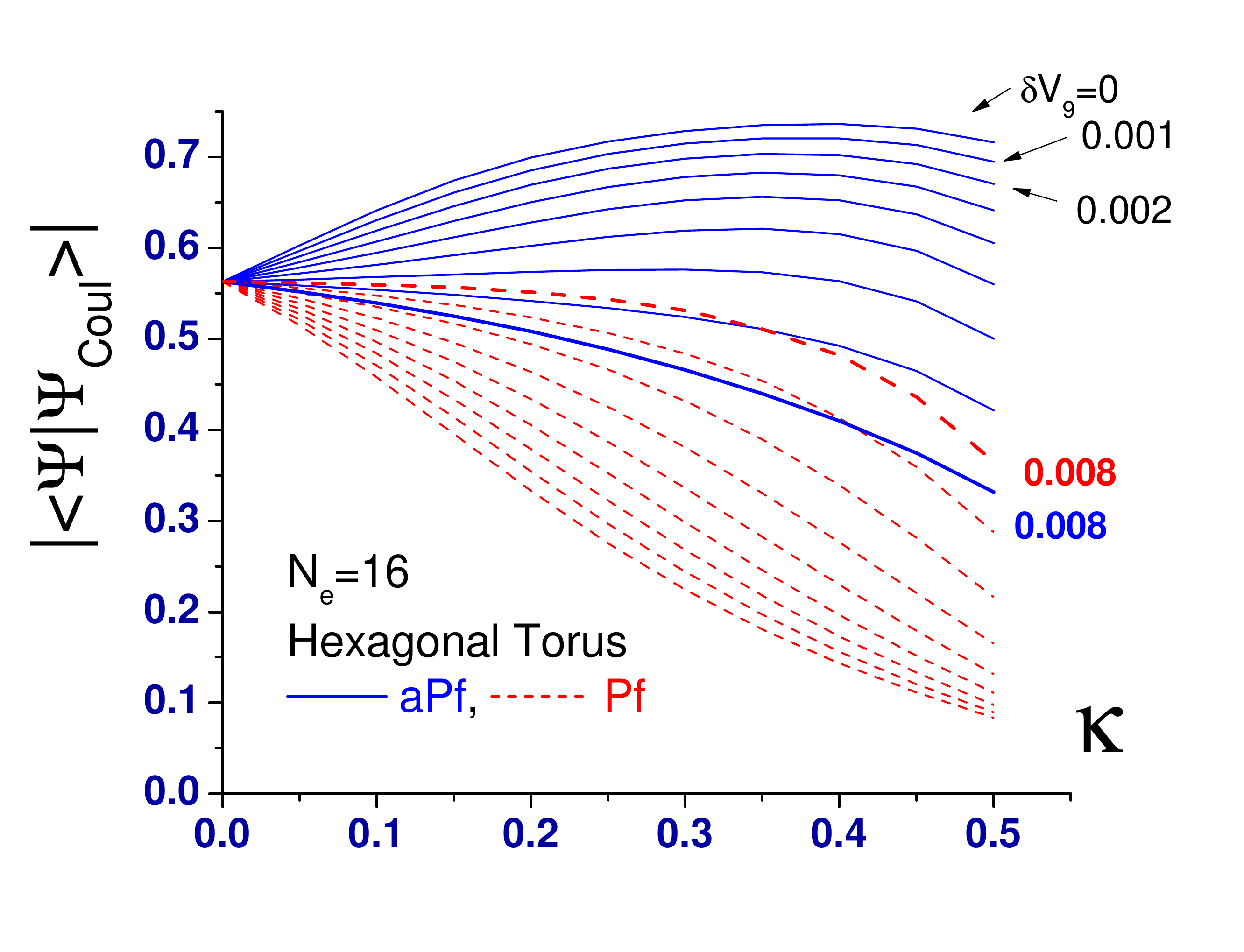}
         \caption{ (Color online) Plot of overlaps for the aPf 
(solid) and the Pf (dashed) as a function of  $\kappa$ for a sequence of
$\delta V_9$ values in increments of 0.001. The top and bottom curves are for $\delta V_9=0$. The 
second top (and bottom) are for 0.001, etc.  The system undergoes a phase transition at 
$\delta V_9\approx 0.008$. 
                        }
        \label{fig:torusovsvskap}
      \end{figure}

Since there is no shift on the torus, comparison of the two states is straightforward. 
Fig. \ref{fig:torusovsvskap} shows the overlaps as a function of
$\kappa$ for a series of $\delta V_9$ values. The dependence on $V_9$ for each $\kappa$ is also linear (not shown).  In contrast to the sphere results for small $\kappa$, 
there is opposing dependence of the overlaps on $\kappa$, with the aPf increasing 
and Pf decreasing.  Increasing $\delta V_9$ past $0.008$ reverses the trends and 
the overlap with the Pf increases with $\kappa$.

{\it Discussion.--} 
Entanglement properties rather than high overlaps are a better indicator of which topological
phase of matter a particular state may belong to. These have already been 
reported for the aPf ground state of the Coulomb potential, both on the cylinder\cite{Zaletel-etal}
and on the sphere\cite{Pakrouski-etal}.  However, very high overlaps can be
reached (97\%, $\kappa=2.1$ for $N_e=14$ on the sphere and 96\%, $\kappa=1.5$ for $N_e=16$ on the hexagonal torus)  by adiabatically varying the effective Hamiltonian to include only three ($V_3$, $V_6$, and $V_9$) of the 3-body pseudo-potentials and the Bishara-Nayak 2-body pseudo-potentials used by 
Woj's {\it et. al.}  This can be done  without  encountering a phase transition.  While these parameters may seem unjustified or
unphysical they do establish a broader phase diagram for the aPf ground state.

We have shown that under plausible experimental conditions the aPf is found to be 
favored for small LL-mixing parameters. 
 The only exception is when just the 
lowest two LLs are kept, then the Pf is preferred. 
A reversal to the Pf at very large sizes cannot be ruled out by 
the present study. However,
the aPf ground state appears to be robust, particularly on the torus, 
making such a reversal unlikely.

\begin{acknowledgments}
I thank Jainendra Jain, Thierry Jolicoeur, Kyril Pakrouski, and Mike Peterson
for discussions and correspondence. I also thank 
Thierry Jolicoeur, Roger Mong, Zlatko Papic, Steve Simon, and Mike Zaletel for comments and suggestions and 
particularly Mike Zaletel for encouraging me to resolve the discrepancy between the two 
approaches.  This work is supported by DOE Grant DE-SC0002140.
\end{acknowledgments}

\bibliography{ref}

\begin{thebibliography}{42}%
\makeatletter
\providecommand \@ifxundefined [1]{%
 \@ifx{#1\undefined}
}%
\providecommand \@ifnum [1]{%
 \ifnum #1\expandafter \@firstoftwo
 \else \expandafter \@secondoftwo
 \fi
}%
\providecommand \@ifx [1]{%
 \ifx #1\expandafter \@firstoftwo
 \else \expandafter \@secondoftwo
 \fi
}%
\providecommand \natexlab [1]{#1}%
\providecommand \enquote  [1]{``#1''}%
\providecommand \bibnamefont  [1]{#1}%
\providecommand \bibfnamefont [1]{#1}%
\providecommand \citenamefont [1]{#1}%
\providecommand \href@noop [0]{\@secondoftwo}%
\providecommand \href [0]{\begingroup \@sanitize@url \@href}%
\providecommand \@href[1]{\@@startlink{#1}\@@href}%
\providecommand \@@href[1]{\endgroup#1\@@endlink}%
\providecommand \@sanitize@url [0]{\catcode `\\12\catcode `\$12\catcode
  `\&12\catcode `\#12\catcode `\^12\catcode `\_12\catcode `\%12\relax}%
\providecommand \@@startlink[1]{}%
\providecommand \@@endlink[0]{}%
\providecommand \url  [0]{\begingroup\@sanitize@url \@url }%
\providecommand \@url [1]{\endgroup\@href {#1}{\urlprefix }}%
\providecommand \urlprefix  [0]{URL }%
\providecommand \Eprint [0]{\href }%
\providecommand \doibase [0]{http://dx.doi.org/}%
\providecommand \selectlanguage [0]{\@gobble}%
\providecommand \bibinfo  [0]{\@secondoftwo}%
\providecommand \bibfield  [0]{\@secondoftwo}%
\providecommand \translation [1]{[#1]}%
\providecommand \BibitemOpen [0]{}%
\providecommand \bibitemStop [0]{}%
\providecommand \bibitemNoStop [0]{.\EOS\space}%
\providecommand \EOS [0]{\spacefactor3000\relax}%
\providecommand \BibitemShut  [1]{\csname bibitem#1\endcsname}%
\let\auto@bib@innerbib\@empty
\bibitem [{\citenamefont {Wen}(1992)}]{Wen-Top-Order}%
  \BibitemOpen
  \bibfield  {author} {\bibinfo {author} {\bibfnamefont {X.-G.}\ \bibnamefont
  {Wen}},\ }\href {\doibase 10.1142/S0217979292000840} {\bibfield  {journal}
  {\bibinfo  {journal} {International Journal of Modern Physics B}\ }\textbf
  {\bibinfo {volume} {06}},\ \bibinfo {pages} {1711} (\bibinfo {year}
  {1992})}\BibitemShut {NoStop}%
\bibitem [{\citenamefont {Tsui}\ and\ \citenamefont
  {Gossard}(1982)}]{Tsui-etal}%
  \BibitemOpen
  \bibfield  {author} {\bibinfo {author} {\bibfnamefont {H.~L.}\ \bibnamefont
  {Tsui}, \bibfnamefont {D.~C.~Stormer}}\ and\ \bibinfo {author} {\bibfnamefont
  {A.~C.}\ \bibnamefont {Gossard}},\ }\href@noop {} {\bibfield  {journal}
  {\bibinfo  {journal} {Phys. Rev. Lett.}\ }\textbf {\bibinfo {volume} {48}},\
  \bibinfo {pages} {1559} (\bibinfo {year} {1982})}\BibitemShut {NoStop}%
\bibitem [{\citenamefont {Laughlin}(1983{\natexlab{a}})}]{Laughlin}%
  \BibitemOpen
  \bibfield  {author} {\bibinfo {author} {\bibfnamefont {R.~B.}\ \bibnamefont
  {Laughlin}},\ }\href@noop {} {\bibfield  {journal} {\bibinfo  {journal}
  {Phys. Rev. Lett.}\ }\textbf {\bibinfo {volume} {50}},\ \bibinfo {pages}
  {1395} (\bibinfo {year} {1983}{\natexlab{a}})}\BibitemShut {NoStop}%
\bibitem [{\citenamefont {Willett}\ \emph {et~al.}(1987)\citenamefont
  {Willett}, \citenamefont {Eisenstein}, \citenamefont {St\"ormer},
  \citenamefont {Tsui}, \citenamefont {Gossard},\ and\ \citenamefont
  {English}}]{Willett-even}%
  \BibitemOpen
  \bibfield  {author} {\bibinfo {author} {\bibfnamefont {R.}~\bibnamefont
  {Willett}}, \bibinfo {author} {\bibfnamefont {J.~P.}\ \bibnamefont
  {Eisenstein}}, \bibinfo {author} {\bibfnamefont {H.~L.}\ \bibnamefont
  {St\"ormer}}, \bibinfo {author} {\bibfnamefont {D.~C.}\ \bibnamefont {Tsui}},
  \bibinfo {author} {\bibfnamefont {A.~C.}\ \bibnamefont {Gossard}}, \ and\
  \bibinfo {author} {\bibfnamefont {J.~H.}\ \bibnamefont {English}},\ }\href
  {\doibase 10.1103/PhysRevLett.59.1776} {\bibfield  {journal} {\bibinfo
  {journal} {Phys. Rev. Lett.}\ }\textbf {\bibinfo {volume} {59}},\ \bibinfo
  {pages} {1776} (\bibinfo {year} {1987})}\BibitemShut {NoStop}%
\bibitem [{\citenamefont {Pan}\ \emph {et~al.}(1999)\citenamefont {Pan},
  \citenamefont {Xia}, \citenamefont {Shvarts}, \citenamefont {Adams},
  \citenamefont {Stormer}, \citenamefont {Tsui}, \citenamefont {Pfeiffer},
  \citenamefont {Baldwin},\ and\ \citenamefont {West}}]{Pan-etal-5-2}%
  \BibitemOpen
  \bibfield  {author} {\bibinfo {author} {\bibfnamefont {W.}~\bibnamefont
  {Pan}}, \bibinfo {author} {\bibfnamefont {J.-S.}\ \bibnamefont {Xia}},
  \bibinfo {author} {\bibfnamefont {V.}~\bibnamefont {Shvarts}}, \bibinfo
  {author} {\bibfnamefont {D.~E.}\ \bibnamefont {Adams}}, \bibinfo {author}
  {\bibfnamefont {H.~L.}\ \bibnamefont {Stormer}}, \bibinfo {author}
  {\bibfnamefont {D.~C.}\ \bibnamefont {Tsui}}, \bibinfo {author}
  {\bibfnamefont {L.~N.}\ \bibnamefont {Pfeiffer}}, \bibinfo {author}
  {\bibfnamefont {K.~W.}\ \bibnamefont {Baldwin}}, \ and\ \bibinfo {author}
  {\bibfnamefont {K.~W.}\ \bibnamefont {West}},\ }\href {\doibase
  10.1103/PhysRevLett.83.3530} {\bibfield  {journal} {\bibinfo  {journal}
  {Phys. Rev. Lett.}\ }\textbf {\bibinfo {volume} {83}},\ \bibinfo {pages}
  {3530} (\bibinfo {year} {1999})}\BibitemShut {NoStop}%
\bibitem [{\citenamefont {Xia}\ \emph {et~al.}(2004)\citenamefont {Xia},
  \citenamefont {Pan}, \citenamefont {Vicente}, \citenamefont {Adams},
  \citenamefont {Sullivan}, \citenamefont {Stormer}, \citenamefont {Tsui},
  \citenamefont {Pfeiffer}, \citenamefont {Baldwin},\ and\ \citenamefont
  {West}}]{Xia-etal}%
  \BibitemOpen
  \bibfield  {author} {\bibinfo {author} {\bibfnamefont {J.~S.}\ \bibnamefont
  {Xia}}, \bibinfo {author} {\bibfnamefont {W.}~\bibnamefont {Pan}}, \bibinfo
  {author} {\bibfnamefont {C.~L.}\ \bibnamefont {Vicente}}, \bibinfo {author}
  {\bibfnamefont {E.~D.}\ \bibnamefont {Adams}}, \bibinfo {author}
  {\bibfnamefont {N.~S.}\ \bibnamefont {Sullivan}}, \bibinfo {author}
  {\bibfnamefont {H.~L.}\ \bibnamefont {Stormer}}, \bibinfo {author}
  {\bibfnamefont {D.~C.}\ \bibnamefont {Tsui}}, \bibinfo {author}
  {\bibfnamefont {L.~N.}\ \bibnamefont {Pfeiffer}}, \bibinfo {author}
  {\bibfnamefont {K.~W.}\ \bibnamefont {Baldwin}}, \ and\ \bibinfo {author}
  {\bibfnamefont {K.~W.}\ \bibnamefont {West}},\ }\href {\doibase
  10.1103/PhysRevLett.93.176809} {\bibfield  {journal} {\bibinfo  {journal}
  {Phys. Rev. Lett.}\ }\textbf {\bibinfo {volume} {93}},\ \bibinfo {pages}
  {176809} (\bibinfo {year} {2004})}\BibitemShut {NoStop}%
\bibitem [{\citenamefont {Moore}\ and\ \citenamefont {Read}(1991)}]{MR}%
  \BibitemOpen
  \bibfield  {author} {\bibinfo {author} {\bibfnamefont {G.}~\bibnamefont
  {Moore}}\ and\ \bibinfo {author} {\bibfnamefont {N.}~\bibnamefont {Read}},\
  }\href {\doibase http://dx.doi.org/10.1016/0550-3213(91)90407-O} {\bibfield
  {journal} {\bibinfo  {journal} {Nuclear Physics B}\ }\textbf {\bibinfo
  {volume} {360}},\ \bibinfo {pages} {362 } (\bibinfo {year}
  {1991})}\BibitemShut {NoStop}%
\bibitem [{\citenamefont {Read}\ and\ \citenamefont
  {Green}(2000)}]{Read-Green}%
  \BibitemOpen
  \bibfield  {author} {\bibinfo {author} {\bibfnamefont {N.}~\bibnamefont
  {Read}}\ and\ \bibinfo {author} {\bibfnamefont {D.}~\bibnamefont {Green}},\
  }\href {\doibase 10.1103/PhysRevB.61.10267} {\bibfield  {journal} {\bibinfo
  {journal} {Phys. Rev. B}\ }\textbf {\bibinfo {volume} {61}},\ \bibinfo
  {pages} {10267} (\bibinfo {year} {2000})}\BibitemShut {NoStop}%
\bibitem [{\citenamefont {Jain}(1989)}]{Jain}%
  \BibitemOpen
  \bibfield  {author} {\bibinfo {author} {\bibfnamefont {J.~K.}\ \bibnamefont
  {Jain}},\ }\href {\doibase 10.1103/PhysRevLett.63.199} {\bibfield  {journal}
  {\bibinfo  {journal} {Phys. Rev. Lett.}\ }\textbf {\bibinfo {volume} {63}},\
  \bibinfo {pages} {199} (\bibinfo {year} {1989})}\BibitemShut {NoStop}%
\bibitem [{\citenamefont {Nayak}\ and\ \citenamefont
  {Wilczek}(1994)}]{Nayak-Wilczek}%
  \BibitemOpen
  \bibfield  {author} {\bibinfo {author} {\bibfnamefont {C.}~\bibnamefont
  {Nayak}}\ and\ \bibinfo {author} {\bibfnamefont {F.}~\bibnamefont
  {Wilczek}},\ }\href {\doibase http://dx.doi.org/10.1016/0550-3213(94)90477-4}
  {\bibfield  {journal} {\bibinfo  {journal} {Nuclear Physics B}\ }\textbf
  {\bibinfo {volume} {417}},\ \bibinfo {pages} {359 } (\bibinfo {year}
  {1994})}\BibitemShut {NoStop}%
\bibitem [{\citenamefont {Read}\ and\ \citenamefont {Rezayi}(1996)}]{RR-MRQH}%
  \BibitemOpen
  \bibfield  {author} {\bibinfo {author} {\bibfnamefont {N.}~\bibnamefont
  {Read}}\ and\ \bibinfo {author} {\bibfnamefont {E.}~\bibnamefont {Rezayi}},\
  }\href {\doibase 10.1103/PhysRevB.54.16864} {\bibfield  {journal} {\bibinfo
  {journal} {Phys. Rev. B}\ }\textbf {\bibinfo {volume} {54}},\ \bibinfo
  {pages} {16864} (\bibinfo {year} {1996})}\BibitemShut {NoStop}%
\bibitem [{\citenamefont {Read}(2009)}]{Read-p+ip}%
  \BibitemOpen
  \bibfield  {author} {\bibinfo {author} {\bibfnamefont {N.}~\bibnamefont
  {Read}},\ }\href {\doibase 10.1103/PhysRevB.79.045308} {\bibfield  {journal}
  {\bibinfo  {journal} {Phys. Rev. B}\ }\textbf {\bibinfo {volume} {79}},\
  \bibinfo {pages} {045308} (\bibinfo {year} {2009})}\BibitemShut {NoStop}%
\bibitem [{\citenamefont {Kitaev}(2006)}]{Kitaev-anyons}%
  \BibitemOpen
  \bibfield  {author} {\bibinfo {author} {\bibfnamefont {A.}~\bibnamefont
  {Kitaev}},\ }\href {\doibase http://dx.doi.org/10.1016/j.aop.2005.10.005}
  {\bibfield  {journal} {\bibinfo  {journal} {Annals of Physics}\ }\textbf
  {\bibinfo {volume} {321}},\ \bibinfo {pages} {2 } (\bibinfo {year} {2006})},\
   \bibinfo {note} {January Special Issue}\BibitemShut {NoStop}%
\bibitem [{\citenamefont {Nayak}\ \emph {et~al.}(2008)\citenamefont {Nayak},
  \citenamefont {Simon}, \citenamefont {Stern}, \citenamefont {Freedman},\ and\
  \citenamefont {Das~Sarma}}]{Nayak-etal-RMP}%
  \BibitemOpen
  \bibfield  {author} {\bibinfo {author} {\bibfnamefont {C.}~\bibnamefont
  {Nayak}}, \bibinfo {author} {\bibfnamefont {S.~H.}\ \bibnamefont {Simon}},
  \bibinfo {author} {\bibfnamefont {A.}~\bibnamefont {Stern}}, \bibinfo
  {author} {\bibfnamefont {M.}~\bibnamefont {Freedman}}, \ and\ \bibinfo
  {author} {\bibfnamefont {S.}~\bibnamefont {Das~Sarma}},\ }\href {\doibase
  10.1103/RevModPhys.80.1083} {\bibfield  {journal} {\bibinfo  {journal} {Rev.
  Mod. Phys.}\ }\textbf {\bibinfo {volume} {80}},\ \bibinfo {pages} {1083}
  (\bibinfo {year} {2008})}\BibitemShut {NoStop}%
\bibitem [{\citenamefont {Morf}(1998)}]{Morf}%
  \BibitemOpen
  \bibfield  {author} {\bibinfo {author} {\bibfnamefont {R.~H.}\ \bibnamefont
  {Morf}},\ }\href {\doibase 10.1103/PhysRevLett.80.1505} {\bibfield  {journal}
  {\bibinfo  {journal} {Phys. Rev. Lett.}\ }\textbf {\bibinfo {volume} {80}},\
  \bibinfo {pages} {1505} (\bibinfo {year} {1998})}\BibitemShut {NoStop}%
\bibitem [{\citenamefont {Rezayi}\ and\ \citenamefont
  {Haldane}(2000)}]{Rezayi-Haldane-5/2}%
  \BibitemOpen
  \bibfield  {author} {\bibinfo {author} {\bibfnamefont {E.~H.}\ \bibnamefont
  {Rezayi}}\ and\ \bibinfo {author} {\bibfnamefont {F.~D.~M.}\ \bibnamefont
  {Haldane}},\ }\href {\doibase 10.1103/PhysRevLett.84.4685} {\bibfield
  {journal} {\bibinfo  {journal} {Phys. Rev. Lett.}\ }\textbf {\bibinfo
  {volume} {84}},\ \bibinfo {pages} {4685} (\bibinfo {year}
  {2000})}\BibitemShut {NoStop}%
\bibitem [{\citenamefont {Wan}\ \emph {et~al.}(2008)\citenamefont {Wan},
  \citenamefont {Hu}, \citenamefont {Rezayi},\ and\ \citenamefont
  {Yang}}]{XinWan-etal}%
  \BibitemOpen
  \bibfield  {author} {\bibinfo {author} {\bibfnamefont {X.}~\bibnamefont
  {Wan}}, \bibinfo {author} {\bibfnamefont {Z.-X.}\ \bibnamefont {Hu}},
  \bibinfo {author} {\bibfnamefont {E.~H.}\ \bibnamefont {Rezayi}}, \ and\
  \bibinfo {author} {\bibfnamefont {K.}~\bibnamefont {Yang}},\ }\href {\doibase
  10.1103/PhysRevB.77.165316} {\bibfield  {journal} {\bibinfo  {journal} {Phys.
  Rev. B}\ }\textbf {\bibinfo {volume} {77}},\ \bibinfo {pages} {165316}
  (\bibinfo {year} {2008})}\BibitemShut {NoStop}%
\bibitem [{\citenamefont {Zhao}\ \emph {et~al.}(2011)\citenamefont {Zhao},
  \citenamefont {Sheng},\ and\ \citenamefont {Haldane}}]{Sheng-Haldane-DMRG}%
  \BibitemOpen
  \bibfield  {author} {\bibinfo {author} {\bibfnamefont {J.}~\bibnamefont
  {Zhao}}, \bibinfo {author} {\bibfnamefont {D.~N.}\ \bibnamefont {Sheng}}, \
  and\ \bibinfo {author} {\bibfnamefont {F.~D.~M.}\ \bibnamefont {Haldane}},\
  }\href {\doibase 10.1103/PhysRevB.83.195135} {\bibfield  {journal} {\bibinfo
  {journal} {Phys. Rev. B}\ }\textbf {\bibinfo {volume} {83}},\ \bibinfo
  {pages} {195135} (\bibinfo {year} {2011})}\BibitemShut {NoStop}%
\bibitem [{\citenamefont {Levin}\ \emph {et~al.}(2007)\citenamefont {Levin},
  \citenamefont {Halperin},\ and\ \citenamefont
  {Rosenow}}]{Levin-Halperin-Rosenow}%
  \BibitemOpen
  \bibfield  {author} {\bibinfo {author} {\bibfnamefont {M.}~\bibnamefont
  {Levin}}, \bibinfo {author} {\bibfnamefont {B.~I.}\ \bibnamefont {Halperin}},
  \ and\ \bibinfo {author} {\bibfnamefont {B.}~\bibnamefont {Rosenow}},\ }\href
  {\doibase 10.1103/PhysRevLett.99.236806} {\bibfield  {journal} {\bibinfo
  {journal} {Phys. Rev. Lett.}\ }\textbf {\bibinfo {volume} {99}},\ \bibinfo
  {pages} {236806} (\bibinfo {year} {2007})}\BibitemShut {NoStop}%
\bibitem [{\citenamefont {Lee}\ \emph {et~al.}(2007)\citenamefont {Lee},
  \citenamefont {Ryu}, \citenamefont {Nayak},\ and\ \citenamefont
  {Fisher}}]{Nayak-Fisher}%
  \BibitemOpen
  \bibfield  {author} {\bibinfo {author} {\bibfnamefont {S.-S.}\ \bibnamefont
  {Lee}}, \bibinfo {author} {\bibfnamefont {S.}~\bibnamefont {Ryu}}, \bibinfo
  {author} {\bibfnamefont {C.}~\bibnamefont {Nayak}}, \ and\ \bibinfo {author}
  {\bibfnamefont {M.~P.~A.}\ \bibnamefont {Fisher}},\ }\href {\doibase
  10.1103/PhysRevLett.99.236807} {\bibfield  {journal} {\bibinfo  {journal}
  {Phys. Rev. Lett.}\ }\textbf {\bibinfo {volume} {99}},\ \bibinfo {pages}
  {236807} (\bibinfo {year} {2007})}\BibitemShut {NoStop}%
\bibitem [{\citenamefont {Rezayi}\ and\ \citenamefont
  {Simon}(2011)}]{Rezayi-Simon-LLmix}%
  \BibitemOpen
  \bibfield  {author} {\bibinfo {author} {\bibfnamefont {E.~H.}\ \bibnamefont
  {Rezayi}}\ and\ \bibinfo {author} {\bibfnamefont {S.~H.}\ \bibnamefont
  {Simon}},\ }\href {\doibase 10.1103/PhysRevLett.106.116801} {\bibfield
  {journal} {\bibinfo  {journal} {Phys. Rev. Lett.}\ }\textbf {\bibinfo
  {volume} {106}},\ \bibinfo {pages} {116801} (\bibinfo {year}
  {2011})}\BibitemShut {NoStop}%
\bibitem [{\citenamefont {Zaletel}\ \emph {et~al.}(2015)\citenamefont
  {Zaletel}, \citenamefont {Mong}, \citenamefont {Pollmann},\ and\
  \citenamefont {Rezayi}}]{Zaletel-etal}%
  \BibitemOpen
  \bibfield  {author} {\bibinfo {author} {\bibfnamefont {M.~P.}\ \bibnamefont
  {Zaletel}}, \bibinfo {author} {\bibfnamefont {R.~S.~K.}\ \bibnamefont
  {Mong}}, \bibinfo {author} {\bibfnamefont {F.}~\bibnamefont {Pollmann}}, \
  and\ \bibinfo {author} {\bibfnamefont {E.~H.}\ \bibnamefont {Rezayi}},\
  }\href {\doibase 10.1103/PhysRevB.91.045115} {\bibfield  {journal} {\bibinfo
  {journal} {Phys. Rev. B}\ }\textbf {\bibinfo {volume} {91}},\ \bibinfo
  {pages} {045115} (\bibinfo {year} {2015})}\BibitemShut {NoStop}%
\bibitem [{\citenamefont {Zaletel}\ \emph {et~al.}(2013)\citenamefont
  {Zaletel}, \citenamefont {Mong},\ and\ \citenamefont
  {Pollmann}}]{Zaletel-Mong-Pollmann}%
  \BibitemOpen
  \bibfield  {author} {\bibinfo {author} {\bibfnamefont {M.~P.}\ \bibnamefont
  {Zaletel}}, \bibinfo {author} {\bibfnamefont {R.~S.~K.}\ \bibnamefont
  {Mong}}, \ and\ \bibinfo {author} {\bibfnamefont {F.}~\bibnamefont
  {Pollmann}},\ }\href {\doibase 10.1103/PhysRevLett.110.236801} {\bibfield
  {journal} {\bibinfo  {journal} {Phys. Rev. Lett.}\ }\textbf {\bibinfo
  {volume} {110}},\ \bibinfo {pages} {236801} (\bibinfo {year}
  {2013})}\BibitemShut {NoStop}%
\bibitem [{\citenamefont {Zaletel}\ and\ \citenamefont
  {Mong}(2012)}]{Zaletel-Mong}%
  \BibitemOpen
  \bibfield  {author} {\bibinfo {author} {\bibfnamefont {M.~P.}\ \bibnamefont
  {Zaletel}}\ and\ \bibinfo {author} {\bibfnamefont {R.~S.~K.}\ \bibnamefont
  {Mong}},\ }\href {\doibase 10.1103/PhysRevB.86.245305} {\bibfield  {journal}
  {\bibinfo  {journal} {Phys. Rev. B}\ }\textbf {\bibinfo {volume} {86}},\
  \bibinfo {pages} {245305} (\bibinfo {year} {2012})}\BibitemShut {NoStop}%
\bibitem [{\citenamefont {Estienne}\ \emph {et~al.}(2013)\citenamefont
  {Estienne}, \citenamefont {Papi\ifmmode~\acute{c}\else \'{c}\fi{}},
  \citenamefont {Regnault},\ and\ \citenamefont {Bernevig}}]{Bernevig-etal}%
  \BibitemOpen
  \bibfield  {author} {\bibinfo {author} {\bibfnamefont {B.}~\bibnamefont
  {Estienne}}, \bibinfo {author} {\bibfnamefont {Z.}~\bibnamefont
  {Papi\ifmmode~\acute{c}\else \'{c}\fi{}}}, \bibinfo {author} {\bibfnamefont
  {N.}~\bibnamefont {Regnault}}, \ and\ \bibinfo {author} {\bibfnamefont
  {B.~A.}\ \bibnamefont {Bernevig}},\ }\href {\doibase
  10.1103/PhysRevB.87.161112} {\bibfield  {journal} {\bibinfo  {journal} {Phys.
  Rev. B}\ }\textbf {\bibinfo {volume} {87}},\ \bibinfo {pages} {161112}
  (\bibinfo {year} {2013})}\BibitemShut {NoStop}%
\bibitem [{\citenamefont {W\'ojs}\ \emph {et~al.}(2010)\citenamefont {W\'ojs},
  \citenamefont {T\ifmmode~\mbox{\H{o}}\else \H{o}\fi{}ke},\ and\ \citenamefont
  {Jain}}]{Wojs-Toke-Jain}%
  \BibitemOpen
  \bibfield  {author} {\bibinfo {author} {\bibfnamefont {A.}~\bibnamefont
  {W\'ojs}}, \bibinfo {author} {\bibfnamefont {C.}~\bibnamefont
  {T\ifmmode~\mbox{\H{o}}\else \H{o}\fi{}ke}}, \ and\ \bibinfo {author}
  {\bibfnamefont {J.~K.}\ \bibnamefont {Jain}},\ }\href {\doibase
  10.1103/PhysRevLett.105.096802} {\bibfield  {journal} {\bibinfo  {journal}
  {Phys. Rev. Lett.}\ }\textbf {\bibinfo {volume} {105}},\ \bibinfo {pages}
  {096802} (\bibinfo {year} {2010})}\BibitemShut {NoStop}%
\bibitem [{\citenamefont {Bishara}\ and\ \citenamefont
  {Nayak}(2009)}]{Bishara-Nayak}%
  \BibitemOpen
  \bibfield  {author} {\bibinfo {author} {\bibfnamefont {W.}~\bibnamefont
  {Bishara}}\ and\ \bibinfo {author} {\bibfnamefont {C.}~\bibnamefont
  {Nayak}},\ }\href {\doibase 10.1103/PhysRevB.80.121302} {\bibfield  {journal}
  {\bibinfo  {journal} {Phys. Rev. B}\ }\textbf {\bibinfo {volume} {80}},\
  \bibinfo {pages} {121302} (\bibinfo {year} {2009})}\BibitemShut {NoStop}%
\bibitem [{\citenamefont {Pakrouski}\ \emph {et~al.}(2015)\citenamefont
  {Pakrouski}, \citenamefont {Peterson}, \citenamefont {Jolicoeur},
  \citenamefont {Scarola}, \citenamefont {Nayak},\ and\ \citenamefont
  {Troyer}}]{Pakrouski-etal}%
  \BibitemOpen
  \bibfield  {author} {\bibinfo {author} {\bibfnamefont {K.}~\bibnamefont
  {Pakrouski}}, \bibinfo {author} {\bibfnamefont {M.~R.}\ \bibnamefont
  {Peterson}}, \bibinfo {author} {\bibfnamefont {T.}~\bibnamefont {Jolicoeur}},
  \bibinfo {author} {\bibfnamefont {V.~W.}\ \bibnamefont {Scarola}}, \bibinfo
  {author} {\bibfnamefont {C.}~\bibnamefont {Nayak}}, \ and\ \bibinfo {author}
  {\bibfnamefont {M.}~\bibnamefont {Troyer}},\ }\href {\doibase
  10.1103/PhysRevX.5.021004} {\bibfield  {journal} {\bibinfo  {journal} {Phys.
  Rev. X}\ }\textbf {\bibinfo {volume} {5}},\ \bibinfo {pages} {021004}
  (\bibinfo {year} {2015})}\BibitemShut {NoStop}%
\bibitem [{\citenamefont {Haldane}(1983)}]{Haldane-2body}%
  \BibitemOpen
  \bibfield  {author} {\bibinfo {author} {\bibfnamefont {F.~D.~M.}\
  \bibnamefont {Haldane}},\ }\href {\doibase 10.1103/PhysRevLett.51.605}
  {\bibfield  {journal} {\bibinfo  {journal} {Phys. Rev. Lett.}\ }\textbf
  {\bibinfo {volume} {51}},\ \bibinfo {pages} {605} (\bibinfo {year}
  {1983})}\BibitemShut {NoStop}%
\bibitem [{\citenamefont {Simon}\ \emph {et~al.}(2007)\citenamefont {Simon},
  \citenamefont {Rezayi},\ and\ \citenamefont {Cooper}}]{Simon-Rezayi-Cooper}%
  \BibitemOpen
  \bibfield  {author} {\bibinfo {author} {\bibfnamefont {S.~H.}\ \bibnamefont
  {Simon}}, \bibinfo {author} {\bibfnamefont {E.~H.}\ \bibnamefont {Rezayi}}, \
  and\ \bibinfo {author} {\bibfnamefont {N.~R.}\ \bibnamefont {Cooper}},\
  }\href {\doibase 10.1103/PhysRevB.75.195306} {\bibfield  {journal} {\bibinfo
  {journal} {Phys. Rev. B}\ }\textbf {\bibinfo {volume} {75}},\ \bibinfo
  {pages} {195306} (\bibinfo {year} {2007})}\BibitemShut {NoStop}%
\bibitem [{\citenamefont {Simon}\ and\ \citenamefont
  {Rezayi}(2013)}]{Simon-Rezayi}%
  \BibitemOpen
  \bibfield  {author} {\bibinfo {author} {\bibfnamefont {S.~H.}\ \bibnamefont
  {Simon}}\ and\ \bibinfo {author} {\bibfnamefont {E.~H.}\ \bibnamefont
  {Rezayi}},\ }\href {\doibase 10.1103/PhysRevB.87.155426} {\bibfield
  {journal} {\bibinfo  {journal} {Phys. Rev. B}\ }\textbf {\bibinfo {volume}
  {87}},\ \bibinfo {pages} {155426} (\bibinfo {year} {2013})}\BibitemShut
  {NoStop}%
\bibitem [{\citenamefont {Peterson}\ and\ \citenamefont
  {Nayak}(2013)}]{Peterson-Nayak}%
  \BibitemOpen
  \bibfield  {author} {\bibinfo {author} {\bibfnamefont {M.~R.}\ \bibnamefont
  {Peterson}}\ and\ \bibinfo {author} {\bibfnamefont {C.}~\bibnamefont
  {Nayak}},\ }\href {\doibase 10.1103/PhysRevB.87.245129} {\bibfield  {journal}
  {\bibinfo  {journal} {Phys. Rev. B}\ }\textbf {\bibinfo {volume} {87}},\
  \bibinfo {pages} {245129} (\bibinfo {year} {2013})}\BibitemShut {NoStop}%
\bibitem [{\citenamefont {Sodemann}\ and\ \citenamefont
  {MacDonald}(2013)}]{Sodemann-MacDonald}%
  \BibitemOpen
  \bibfield  {author} {\bibinfo {author} {\bibfnamefont {I.}~\bibnamefont
  {Sodemann}}\ and\ \bibinfo {author} {\bibfnamefont {A.~H.}\ \bibnamefont
  {MacDonald}},\ }\href {\doibase 10.1103/PhysRevB.87.245425} {\bibfield
  {journal} {\bibinfo  {journal} {Phys. Rev. B}\ }\textbf {\bibinfo {volume}
  {87}},\ \bibinfo {pages} {245425} (\bibinfo {year} {2013})}\BibitemShut
  {NoStop}%
\bibitem [{Note1()}]{Note1}%
  \BibitemOpen
  \bibinfo {note} {The shift S is defined by the flux-charge equation $N_\phi
  =\nu _v^{-1} N_e-S$, where $\nu _v$ is the filling factor of the valence
  electrons. On the sphere S=3 and -1 for the Pf and aPf
  respectively.}\BibitemShut {Stop}%
\bibitem [{\citenamefont {Papi\ifmmode~\acute{c}\else \'{c}\fi{}}\ \emph
  {et~al.}(2012)\citenamefont {Papi\ifmmode~\acute{c}\else \'{c}\fi{}},
  \citenamefont {Haldane},\ and\ \citenamefont {Rezayi}}]{Zlatko-QW}%
  \BibitemOpen
  \bibfield  {author} {\bibinfo {author} {\bibfnamefont {Z.}~\bibnamefont
  {Papi\ifmmode~\acute{c}\else \'{c}\fi{}}}, \bibinfo {author} {\bibfnamefont
  {F.~D.~M.}\ \bibnamefont {Haldane}}, \ and\ \bibinfo {author} {\bibfnamefont
  {E.~H.}\ \bibnamefont {Rezayi}},\ }\href {\doibase
  10.1103/PhysRevLett.109.266806} {\bibfield  {journal} {\bibinfo  {journal}
  {Phys. Rev. Lett.}\ }\textbf {\bibinfo {volume} {109}},\ \bibinfo {pages}
  {266806} (\bibinfo {year} {2012})}\BibitemShut {NoStop}%
\bibitem [{\citenamefont {Tinkham}(1964)}]{Tinkham}%
  \BibitemOpen
  \bibfield  {author} {\bibinfo {author} {\bibfnamefont {M.}~\bibnamefont
  {Tinkham}},\ }\href@noop {} {\emph {\bibinfo {title} {Group Theory and
  Quantum Mechanics}}}\ (\bibinfo  {publisher} {McGraw-Hill Book Company},\
  \bibinfo {address} {New York},\ \bibinfo {year} {1964})\BibitemShut {NoStop}%
\bibitem [{\citenamefont {{Eisenstein}}\ \emph {et~al.}(2017)\citenamefont
  {{Eisenstein}}, \citenamefont {{Pfeiffer}},\ and\ \citenamefont
  {{West}}}]{Eisenstein-spin-diode}%
  \BibitemOpen
  \bibfield  {author} {\bibinfo {author} {\bibfnamefont {J.~P.}\ \bibnamefont
  {{Eisenstein}}}, \bibinfo {author} {\bibfnamefont {L.~N.}\ \bibnamefont
  {{Pfeiffer}}}, \ and\ \bibinfo {author} {\bibfnamefont {K.~W.}\ \bibnamefont
  {{West}}},\ }\href@noop {} {\bibfield  {journal} {\bibinfo  {journal} {ArXiv
  e-prints}\ } (\bibinfo {year} {2017})},\ \Eprint
  {http://arxiv.org/abs/1701.04463} {arXiv:1701.04463 [cond-mat.mes-hall]}
  \BibitemShut {NoStop}%
\bibitem [{\citenamefont {Tiemann}\ \emph {et~al.}(2012)\citenamefont
  {Tiemann}, \citenamefont {Gamez}, \citenamefont {Kumada},\ and\ \citenamefont
  {Muraki}}]{Gamez-etal}%
  \BibitemOpen
  \bibfield  {author} {\bibinfo {author} {\bibfnamefont {L.}~\bibnamefont
  {Tiemann}}, \bibinfo {author} {\bibfnamefont {G.}~\bibnamefont {Gamez}},
  \bibinfo {author} {\bibfnamefont {N.}~\bibnamefont {Kumada}}, \ and\ \bibinfo
  {author} {\bibfnamefont {K.}~\bibnamefont {Muraki}},\ }\href {\doibase
  10.1126/science.1216697} {\bibfield  {journal} {\bibinfo  {journal}
  {Science}\ }\textbf {\bibinfo {volume} {335}},\ \bibinfo {pages} {828}
  (\bibinfo {year} {2012})}\BibitemShut {NoStop}%
\bibitem [{\citenamefont {Feiguin}\ \emph {et~al.}(2009)\citenamefont
  {Feiguin}, \citenamefont {Rezayi}, \citenamefont {Yang}, \citenamefont
  {Nayak},\ and\ \citenamefont {Das~Sarma}}]{Feiguin-etal}%
  \BibitemOpen
  \bibfield  {author} {\bibinfo {author} {\bibfnamefont {A.~E.}\ \bibnamefont
  {Feiguin}}, \bibinfo {author} {\bibfnamefont {E.}~\bibnamefont {Rezayi}},
  \bibinfo {author} {\bibfnamefont {K.}~\bibnamefont {Yang}}, \bibinfo {author}
  {\bibfnamefont {C.}~\bibnamefont {Nayak}}, \ and\ \bibinfo {author}
  {\bibfnamefont {S.}~\bibnamefont {Das~Sarma}},\ }\href {\doibase
  10.1103/PhysRevB.79.115322} {\bibfield  {journal} {\bibinfo  {journal} {Phys.
  Rev. B}\ }\textbf {\bibinfo {volume} {79}},\ \bibinfo {pages} {115322}
  (\bibinfo {year} {2009})}\BibitemShut {NoStop}%
\bibitem [{\citenamefont {Landau}\ and\ \citenamefont
  {Lifshitz}(1977)}]{Lan-Lif}%
  \BibitemOpen
  \bibfield  {author} {\bibinfo {author} {\bibfnamefont {L.~D.}\ \bibnamefont
  {Landau}}\ and\ \bibinfo {author} {\bibfnamefont {E.~M.}\ \bibnamefont
  {Lifshitz}},\ }\href@noop {} {\emph {\bibinfo {title} {Quantum Mechanics}}}\
  (\bibinfo  {publisher} {Pergamon Press},\ \bibinfo {address} {Oxford},\
  \bibinfo {year} {1977})\BibitemShut {NoStop}%
\bibitem [{Note2()}]{Note2}%
  \BibitemOpen
  \bibinfo {note} {It can be seen that finite-size effects become larger as $M$
  is increased. For larger $M$s the average relative distance between the
  particles increases and may become comparable to the linear dimension of the
  system. As a result the $M=12$ may be off by 15-20\% of its infinite-size
  value}\BibitemShut {NoStop}%
\bibitem [{\citenamefont {Laughlin}(1983{\natexlab{b}})}]{Laughlin-3part}%
  \BibitemOpen
  \bibfield  {author} {\bibinfo {author} {\bibfnamefont {R.~B.}\ \bibnamefont
  {Laughlin}},\ }\href {\doibase 10.1103/PhysRevB.27.3383} {\bibfield
  {journal} {\bibinfo  {journal} {Phys. Rev. B}\ }\textbf {\bibinfo {volume}
  {27}},\ \bibinfo {pages} {3383} (\bibinfo {year}
  {1983}{\natexlab{b}})}\BibitemShut {NoStop}%
\end{thebibliography}%

\end{document}